# Flat band carrier confinement in magic-angle twisted bilayer graphene

Nikhil Tilak[1], Xinyuan Lai[1], Shuang Wu[1], Zhenyuan Zhang[1],
Mingyu Xu[2,3], Raquel de Almeida Ribeiro[2,3], Paul C Canfield[2,3]
and Eva Y. Andrei[1*]

1 Department of Physics and Astronomy, Rutgers, The State University of New Jersey, 136
Frelinghuysen Rd, Piscataway, NJ 08854

2 Ames Laboratory, U.S. Department of Energy, Ames, Iowa 50011, USA

3 Department of Physics and Astronomy, Iowa State University, Ames, Iowa 50011, USA.

Corresponding author email : eandrei@physics.rutgers.edu

## Abstract

Magic angle twisted bilayer graphene has emerged as a powerful platform for studying strongly correlated electron physics, owing to its almost dispersionless low-energy bands and the ability to tune the band filling by electrostatic gating. Techniques to control the twist angle between graphene layers have led to rapid experimental progress but improving sample quality is essential for separating the delicate correlated-electron physics from disorder effects. Owing to the 2D nature of the system and the relatively low carrier density, the samples are highly susceptible to small doping inhomogeneity which can drastically modify the local potential landscape. This potential disorder is distinct from the twist-angle variation which has been studied elsewhere. Here, by using low temperature scanning tunneling spectroscopy and planar tunneling junction measurements, we demonstrate that flat bands in twisted bilayer graphene can amplify small doping inhomogeneity that surprisingly leads to carrier confinement, which in graphene could previously only be realized in the presence of a strong magnetic field.

# Introduction

When two sheets of Graphene are superposed with a relative twist angle, they develop a moiré superstructure with an angle dependent wavelength which reflects the local stacking variation of the two crystal lattices. Hybridization between the two sets of electronic bands produces a strongly modified, twist-angle dependent band structure[1,2]. Close to a 'magic' angle ($\sim 1°$), this leads to very narrow, almost dispersionless (flat) low energy bands[3,4]. These flat bands are isolated from the dispersive bands by single particle bandgaps[5]. Since the kinetic energy of the electrons in the flat bands is quenched, e-e interactions become important and give rise to phenomena such as correlated insulating states[6], superconductivity[7], emergent ferromagnetism[8,9] etc. The electronic properties of twisted Bilayer Graphene (TBG) were initially explored[1] in naturally occurring TBG synthesized by chemical vapor deposition, through the use of Scanning Tunneling Microscopy/Spectroscopy (STM/STS). More recently, the development of techniques[10] to control the twist angle and the doping level, has expanded the range of capabilities to include global measurements such as magnetotransport [6,7,11-14], electronic compressibility[15], angle-resolved photoemission spectroscopy (ARPES)[16], as well as gated local probes including STM/STS[1,17-20], nano-SQUID-on-tip microscopy[21], local compressibility[22].

Obstacles to progress in this field include twist angle and doping inhomogeneity. Whereas efforts to address the former are being undertaken, the effects of local doping variation have thus far been ignored. Naively, this may be justified by the fact that spatial doping variations[7,11,21] in samples using hexagonal boron nitride (hBN) substrates[23] can be as low as $10^{10} \ cm^{-2}$, which is two orders of magnitude lower than the typical charge density in magic angle TBG, $\sim 10^{12} \ cm^{-2}$. However, as we demonstrate below, even such low levels of density inhomogeneity can radically change the response and electronic properties of the system, obscuring the moiré physics when the Fermi level is aligned to the edge of the flat band.

Using tunneling experiments with a traditional STM as well as a novel planar tunneling device, we find that near the edges of the flat bands, the local doping variations which are ubiquitous in TBG devices, produce patches of conducting regions separated by insulating regions. This leads to carrier confinement on a scale typically larger than the moiré wavelength which can conceal the magic-angle physics.

# Results

We begin by discussing the results of tunneling measurements on a TBG device fabricated by a tear-and-stack technique[17] (see methods for details). The schematic experimental setup is shown in Fig. 1a. We navigate to the micron size sample using a capacitance-based technique[24]. We identify a magic-angle region via STM topography and spectroscopy measurements. The honeycomb lattice of the carbon atoms in graphene is composed of two triangular sublattices labelled A and B. The flat bands are localized mostly on the AA stacking regions, where every atom in the top layer is positioned directly on top of an atom from the bottom layer. These appear as circular bright spots in the topography image when the flat bands are occupied[1] (Fig. 1b). Surrounding the AA regions are six darker regions called AB/BA. In the AB (Bernal stacked) regions, top layer A sublattice atoms are positioned directly above bottom B sublattice atoms while the top layer B sublattice atoms have no partners in the bottom layer. The BA regions are defined similarly via sublattice symmetry. These stacking arrangements are illustrated in Fig. 1c.

The local twist angle($\theta$) is determined by measuring the average moiré wavelength ($L_M$) in the three crystallographic direction using the relation $L_M = a / ((2\sin(\theta/2))$ (1), where $a = 0.246$ nm is the lattice constant of monolayer Graphene[3]. The data presented below were collected in a region with a twist angle of 1.12°, over at least 30 moiré unit cells. The heterostrain in the region was estimated to be 0.2% following reference[18].

Each flat band in TBG is four-fold degenerate owing to the valley and spin degrees of freedom. In total it takes a carrier density of 8 electrons per moiré cell to completely fill both the flat bands. For a twist angle $\theta$ this corresponds to a carrier density $2n_S \approx \frac{16}{a^2\sqrt{3}} \theta^2$ (2) where $\theta$ is measured in radians. We can electrostatically control the carrier density by tuning the voltage ($V_g$) applied between the silicon backgate and the sample bias ($V_b$). The carrier density depends on $V_g$ as, $n = \frac{1}{e}(V_g - V_{g0})\left(\frac{d_1}{\epsilon_0\epsilon_1} + \frac{d_2}{\epsilon_0\epsilon_2}\right)^{-1}$ (3). Here $V_{g0}$ is the gate voltage needed to tune the system to charge neutrality, $\epsilon_0$ is the permittivity of vacuum, e is the electronic charge, $d_i$ and $\epsilon_i$ are the thickness and dielectric constant where $i = 1,2$ correspond to hBN and $SiO_2$ respectively. Given the geometry of our device and the measured twist angle, it takes 75-85 V of backgate to completely fill the empty flat bands.

Figure 1d shows a typical STS curve, which provides a measure of the local density of state (LDOS) within the AA regions when the flat bands are completely filled. The two sharp peaks in the LDOS correspond to two Van Hove singularities (VHS) in the electronic spectrum where the density of states diverges. The rapid decrease in the LDOS near the band edges suggests that close to the empty or full band the electronic properties are particularly susceptible to local doping variations, caused by twist-angle inhomogeneity, impurities or defects. We measure the full width at half-maximum (FWHM) of the electron and hole side flat bands to be $\sim 16$ meV and the two VHS are separated by $\sim 18$ meV. The dips in the $dI/dV_b$ between the flat bands and the remote bands correspond to the single-particle superlattice gaps[5].

Next, we measured the gate voltage dependence of the STS at an AA site (Fig. 2a). In addition to the flat band and the dispersive bands, a series of sharp peaks were observed in the $dI/dV_b$ spectra when the Fermi level was tuned close to the full filling of the electron-side flat band (Fig. 2c). These peaks, which are almost equidistant( $\Delta V_b \approx 60$ meV) in bias voltage, move towards higher bias values as the flat bands are filled. This is opposite to the expected gating behavior for normal features in the DOS which should evolve towards smaller bias values as the bands are filled.

Furthermore, when the peaks intersect the flat band at the Fermi level ($V_b = 0$ mV), a series of diamond like structures appear in the gate dependence map which resembles Coulomb diamonds seen in quantum dots[25]. These can be seen more clearly in the zoomed-in view of the gate dependence map as shown in Fig. 2b. After eliminating trivial explanations of the origin of these peaks (such as dirt stuck to the tip), we concluded that the effect is intrinsic to the sample and not a tip artifact (see Supplementary Note 8). Recalling that, owing to the chiral nature of the quasiparticles, backscattering in graphene is suppressed, the observation of Coulomb diamonds may be surprising. In fact, Coulomb diamonds in graphene are not observed without applying a strong out-of-plane magnetic field. The Magnetic field splits the bands into flat Landau levels separated from each other by an energy gap which depends on the magnitude of the applied Magnetic field[26,27]. Split gates[28] or tip induced band bending[29,30] can be used to locally bend the Landau levels which produce quantum dots in such samples, which are observed as Coulomb diamonds[31] in conductance vs gate voltage maps.

We show that similar to the Landau levels in graphene, the flat bands in magic angle TBG amplify local density variations, but without the need of applying a magnetic field. To this end, we determined the gate voltage at which the flat bands are almost completely full, so that the Fermi level lies at the edge of the flat bands. At this gate voltage, 68 V, we collected spatial $dI/dV_b$ maps, shown in Fig. 3a for bias voltage $V_b = 0$ mV. These maps reveal prominent contrasting patches of bright and dark regions, corresponding to charge puddles created by the local doping inhomogeneity. A comparison of spectra (Fig. 3b) gathered in the bright and dark regions at positions marked by the green and red crosses respectively, shows the flat band is completely full in the dark regions while it is partially empty in the bright regions, directly confirming the spatial doping variation in this sample.

To further characterize this doping variation, we plot in Fig. 3c a line-cut across the conducting region in Fig. 3a along a path denoted by the yellow arrow. The energy of the charge neutrality point ($E_{CNP}$) is marked by the solid black line as a guide to the eye. In the absence of any doping variation, one would expect the $E_{CNP}$ to be constant regardless of the position for a given filling. The black line traces the shape of the potential-well created by the doping disorder at this gate voltage. Since there are no electronic states available near $E_F$ in the regions immediately surrounding the bright islands, the carriers in the bright regions should be confined within them. They can only occupy discrete energy levels whose separation depends on the shape and size of the potential well. Carriers from the tip can tunnel into the bright regions only into these discrete energy levels. These confined carriers can tunnel from the bright conducting islands to the nearby islands through the short insulating barrier separating them. We illustrate this situation with a tunneling diagram in Fig. 3d. Hereafter we refer to these conducting regions as quantum dots.

Valuable information about quantum dots can be extracted by analyzing their charging characteristics. Upon analyzing the coulomb diamonds in Fig. 2b, the capacitances of the quantum dot to the tip ($C_d$), backgate ($C_g$) and the surrounding conducting regions ($C_s$) were found to be in the ratio $C_g : C_s : C_d :: 1 : 37 : 36$ (see Supplementary Note 3). This implies that although $V_b$ is typically small compared to $V_g$, the tip is very efficient at gating the dot because of its larger capacitive coupling. This explains the positive slope of the oblique charging lines in the gate dependence map (Fig. 2a). The size of the quantum dot estimated by using a disc model ($61 \pm 4.8$ nm) is consistent with the size of the conducting island shown in Fig. 3a.

We also performed high resolution $dI/dV_b$ spatial mapping in the conducting region indicated by the dotted magenta square in Fig. 3a in order to directly image the confined carriers (Fig. 3e). The three bright wavefronts in Fig. 3e correspond to three lowest energy states in the conducting region. See the Supplementary Note 5 for more detailed analysis of the evolution of these wavefronts.

Next, we comment on the possible reasons behind the observed doping variation. The TBG rests on a thin hBN flake (23.5 nm). The hBN is exfoliated on top of a commercial highly p-doped Silicon chip with a thermally grown oxide layer (WaferPro, LLC). Silicon dioxide grown on Silicon wafers is known to suffer from dangling bonds, surface defects and charged metallic impurities. In devices where Graphene is directly supported by the Silicon dioxide substrate, these imperfections give rise to charge puddles and to increased scattering[32-35]. The addition of Chlorine gas or Chlorine based hydrocarbons during the growth of the Silicon dioxide on top of Silicon wafers has long been used in the manufacture of Silicon based Metal-Oxide-Semiconductor Field Effect Transistors (MOSFET's) as a way to passivate these metallic impurities[36,37]. The Silicon wafer used in this device, however, was not passivated in this way.

The hBN, which is a wide gap dielectric, serves as a spacer which reduces the effect of the random impurity potential of the $SiO_2$ substrate[2,23,38]. Nonetheless, if the hBN is thin, like in our device (23.5 nm), some of the substrate-induced inhomogeneity can survive. Each of these charged impurities can act as a local gate and change the local charge density of the sample. Furthermore, hBN crystals themselves can host impurities and interstitial defects[39,40] which can dope the sample locally. Yet another possibility is trapped water or organic impurities which inevitably arise during sample fabrication.

Doping inhomogeneity caused by substrate disorder becomes especially important when the bulk Fermi level is tuned close to a band gap, such as the superlattice gaps at the edges of the flat bands or the gaps between Landau levels in a magnetic field. Once the flat bands are doped to integer fillings additional band gaps associated with correlation effects appear at the Fermi level. One should expect to also see quantum confinement near each of these fillings. However, in our STM measurements we did not see a clear signature of confinement when the Fermi level was tuned to integer fillings inside the flat band. There can be two reasons behind this observation: 1. As the Fermi level is tuned deeper inside the flat bands, the size of the quantum dot increases leading to decreased energy spacing of the confined states. This can we washed out by thermal excitations at the temperature of our experiment (4.9 K). Additionally, as the overall density of states is also greater inside the flat band, the disorder potential can be more effectively screened, leading to poorer confinement. 2. Another possibility is that the correlation gaps aren't fully developed at the temperature of the experiment or that the gaps are smaller than our energy resolution. Additional experiments at much lower temperatures should be able to resolve these confinement effects.

To confirm our findings, we studied another TBG device, this time with a novel planar tunneling geometry (Fig. 4a) at 4.2 K and 0.3 K. This device consists of an hBN encapsulated TBG with a local metallic backgate. The thick bottom hBN acts as a gate dielectric while the top hBN is ultra-thin (< 4 layers) and acts as a tunneling barrier. After locating a clean TBG region with an AFM, metallic tunneling electrode is deposited on top of the thin hBN layer through a 600 nm diameter circular hole etched in a thick hBN (see methods). A bias voltage is applied between the tunneling electrode and the TBG and the resulting tunneling current is amplified and measured. Similar to the STM measurement, we obtain the differential conductance ($dI/dV_b$) using a standard Lock-in technique and tune the Fermi level of the sample by applying a gate voltage.

We estimate the twist angle from the measured energy separation of the VHS peaks[1]: $\Delta E_{VHS} \approx \left(\frac{2hv_F}{3a}\right)\theta - 2\,w_\perp$ (4) where $h$ is Plank's constant, $v_F$ is the Fermi velocity of electrons in Graphene, $a$ is the lattice constant of Graphene and $w_\perp \approx 110\ meV$ [3] is the interlayer tunneling parameter. A typical $dI/dV_b$ spectrum of TBG at $V_g = 0$ V is shown in Fig. 4b. Two peaks corresponding to the VHS which flank the charge neutrality point (CNP) are consistent with STS results on TBG close to the magic angle. The measured $\Delta E_{VHS}$ of $48 \pm 2$ meV yields a twist angle of $(1.25 \pm 0.05)°$.

Assuming that the work function of TBG is close to the work function for Bernal stacked bilayer Graphene (4.7 V[41]) and that the work function of the tunneling electrode (Cr/Au) is close to that of Chromium ~ 4.5 eV[42], we can estimate a work function difference of the order of 100 meV between the TBG and the tunneling electrode. This work function difference results in the TBG becoming electron doped and creates a potential-well, approximately the size of the tunneling electrode, where charge carriers can be confined. Similar results were seen previously in STM experiments on Bernal Bilayer graphene[43] where a potential well was created by charging the impurities in hBN. This situation closely resembles the potential-wells observed in the STM device, albeit with different length scales and origins of the confinement potentials. The schematic diagram (Fig. 4d) illustrates this scenario. Indeed, gate-dependence maps measured at 4.2 K (Fig. 4c) show Coulomb diamond like features. On further cooling the sample to 0.3 K (Fig. 4e) and performing high resolution spectroscopic measurements, the resulting gate dependence shows well-developed Coulomb diamonds reflecting the carrier confinement in the device. The charging energy ($E_c$) extracted from the diamond-like features in Fig. 4e, ~2 meV, is much smaller than that observed in the STM device, consistent with a much larger dot size. The diameter of the QD in this case, estimated by using a 2D disc capacitance model[25] (~640 nm), agrees well with the size of the tip-electrode. This supports the scenario of electron confinement due to the strong amplification of potential inhomogeneities in the flat band of magic-angle TBG.

# Discussion

We studied magic-angle TBG devices with STM/STS as well as in a planar tunneling geometry. Sharp peaks in the $dI/dV_b$ spectra near the full filling of the electron-side flat band and their evolution with gate voltage reveals carrier confinement in quantum dots generated by the substrate impurity potential or by the fabrication imposed potential well.

This work demonstrates that, owing to the unique ability of flat bands to amplify small doping inhomogeneities, magic angle TBG samples are susceptible to small random disorder potentials, leading to a drastically modified charge density landscape. These effects, which are not directly observable in non-local measurements such as transport, have thus far been largely overlooked. However, as we have shown, they can nevertheless cause charging inhomogeneity revealing a hitherto hidden source of sample-to sample variation that may contribute to the disparate findings in magic angle TBG. Importantly, the flat-band enabled confinement of charge carriers in graphene demonstrated here, provides a design pathway for achieving on-off switching of graphene devices, which could not be otherwise attained.

# Methods

## STM device

The TBG device was fabricated using the "twist and stack" method. Poly Vinyl Alcohol (PVA) was spin coated on a Silicon chip at 600 RPM followed by a 5 min bake on a hotplate at 90 C. Poly (methyl methacrylate) (PMMA) A-11 (Kayaku Advanced Materials) was then spin coated at 3000 RPM on top of the PVA followed by a 30 minute bake at 90 C. Graphene was exfoliated using tape onto the PMMA/PVA coated chip. The PMMA/PVA stack was then peeled off from the Silicon chip using a piece of tape with a small hole in it. The PVA backing layer was carefully removed with tweezers while leaving the PMMA membrane suspended across the hole in the tape window. If a suitable Graphene flake was found on the PMMA membrane using an optical microscope, a small piece of Polydimethylsiloxane (PDMS) 1 mm thick was place on the PMMA on the back side of the Graphene. This assembly was then transferred to a glass slide creating a glass-slide/PDMS/PMMA/Graphene handle. An hBN flake was separately exfoliated on a highly p-doped Silicon wafer capped by a 285 nm thick thermally grown (non-chlorinated) Silicon Oxide layer (Wafer pro). The glass-slide handle was attached to micromanipulators under an optical microscope and half of the Graphene flake was brought into contact with the hBN flake. When the handle was lifted, the graphene in contact with the hBN was left on the hBN due to strong van der Waals interaction while the other half remained on the PMMA membrane. The chip with the hBN was then rotated by the desired twist angle using a rotation stage and the other half of the graphene was peeled-off onto the first half creating an exposed TBG supported by a thin flake of hBN.

Electrical contact was made to the Graphene with metallic (Au/Ti) electrodes deposited using standard electron beam lithography techniques. A bias voltage was applied to the sample while the tip was grounded. The highly doped Silicon was used as a back gate and enabled us to electrostatically tune the carrier density in the device. The measurements were conducted at a temperature of 4.9 K in a home built STM[24,26,27] using chemically etched Tungsten tips which were tested on the Gold electrode prior to scanning. Tunneling spectra were gathered in constant height mode using a lock-in detection technique by adding a 1-2 mV AC excitation to the DC bias at a frequency of 613 Hz. The sample was annealed in forming gas for 18 h at 270 C before loading into the STM.

The LDOS maps are collected by a Nanonis software routine which measures the topography and $dI/dV_b$ spectrum in the same pass. In brief, at every position in the grid, the tip height is adjusted by the feedback loop based on an initial current and bias setpoint which is the same for all points in the grid. This height is recorded as the topography signal. Then, the feedback loop is disabled while keeping the tip height fixed and the point spectrum is measured by sweeping the sample bias. Thus, each spectrum is measured in a constant-height mode but the height itself depends on the STM topography.

## Planar Tunneling Device

The planar tunneling device consists of two parts, a top hBN layer containing etched holes plus an ultra-thin hBN tunneling layer(t-hBN), and a TBG/hBN stack with a local metal gate.

### Preparation of top hBN with etched holes

Patterning of exfoliated hBN on SiO$_2$ was carried out using e-beam lithography. Then the patterned holes were etched out though CHF3/O2 plasma followed by a lift-off process.

## TBG/hBN stack with a local metal gate

A PVA (poly vinyl alcohol) was spin coated on the silicon chip and baked at 80 C for 10mins, followed by PMMA (Poly methyl-methacrylate) spin-coating and baking at 80 C for 30 mins. Then graphene was exfoliated on the PMMA/PVA/silicon chip. The PMMA thin film was then peeled off from the PVA through the scotch tape with a pre-cut window for subsequent transfer on a bottom hBN flake. The bottom hBN flake is transferred in advance onto a gold electrode as local gate and is annealed at 250 C in Ar/$H_2$ for 6 hours for surface cleaning. Next, part of the target graphene flake on PMMA film was brought in contact with the bottom hBN surface. Due to the stronger van der Waals interaction between graphene and hBN, the part of graphene flake in contact with the hBN could be detached from the PMMA upon lifting the film. Subsequently the substrate with G/hBN/gold stack was rotated by 1°−1.2°. Then the rest part of the graphene flake on PMMA film was brought in contact with the G/hBN surface. The TBG/hBN/gold stack was annealed in forming gas (10% $H_2$, 90% Ar) at 235 C for 6 hours for better adhesion.

## Assembly of the hBN/t-hBN/TBG/gold stacks

The hBN/t-hBN/TBG/hBN/gold stacks are assembled with the dry transfer method in a glovebox (Argon atmosphere), using a stamp consisting of polypropylene carbonate (PPC) film and polydimethylsiloxane (PDMS). The hBN flake with etched holes (30-50nm thick) is firstly picked up. Then the t-hBN (< 4 layers) is picked up by the hBN on the stamp. The hBN/t-hBN stack is then deposited onto the TBG/hBN/gold stack that is prepared separately in step (2). During the assembly of the stack the temperature is kept below 160 C. Atomic force microscopy (AFM) and electrostatic force microscopy (EFM) are subsequently used to identify the region of the TBG[44] prior to depositing the electrical contacts (Cr/Au) for tunneling measurements. The tunneling measurements are carried out in He-3 system with a base temperature of 0.3K. The $dI/dV_b$ measurements are using standard lock-in technique by adding AC excitation of 0.2-1 mV to the DC bias at 7.1 Hz.

# Growth of crystalline hBN

Single crystalline hBN was grown at high temperature and high pressure using a Rockland Research, cubic-multi-anvil press system. Elemental Mg and [11]B, in a Mg:[11]B 1:0.7 molar ratio, were placed in a 7 mm long, 5.7 mm inner diameter BN crucible with any empty space filled by extra BN powder. The crucible was then taken to 3.3 GPa at room temperature and heated to 1380 ℃ over 2 hours. After dwelling at 1380 ℃ for 1 hour, the system was cooled to 580 ℃ over 6 hours and then quenched to room temperature. Once at room temperature the pressure was brought back to ambient conditions. Post growth, the crucible was removed from the pressure media and sealed in an amorphous silica ampoule. Excess Mg was distilled from the hBN by heating one end of the ampoule to 750 ℃ while the other end hung out of the end of a horizontal tube furnace for 200 minutes. Thin, clear flakes of hBN with planar dimensions of up to 1x2 mm$^2$ would be separated from the growth crucible.

## Acknowledgements

NT and XL acknowledge support from the U.S. DOE-BES grant DOE-FG02-99ER45742. S.W and EYA acknowledge support from the Gordon and Betty Moore Foundation's EPiQS initiative grant GBMF9453. MYX and PCC were supported by the U.S. DOE-BES Division of MSE and the research was performed at the Ames Laboratory which is operated for the U.S. DOE by ISU under Contract No. DE-AC02-07CH11358. RAR was supported by the Gordon and Betty Moore Foundation's  EPiQS initiative grant GBMF4411.

## Author contributions

XL and NT fabricated and characterized the STM device. SW and ZZ fabricated and characterized the planar tunneling device. RAR and PCC grew the hBN crystal used to make the devices. NT and EYA wrote the paper with inputs from all authors. EYA supervised the research.

## Competing interests

The authors declare no competing interests.

## Data availability

The data that support the findings of this study are available from the corresponding authors upon reasonable request.

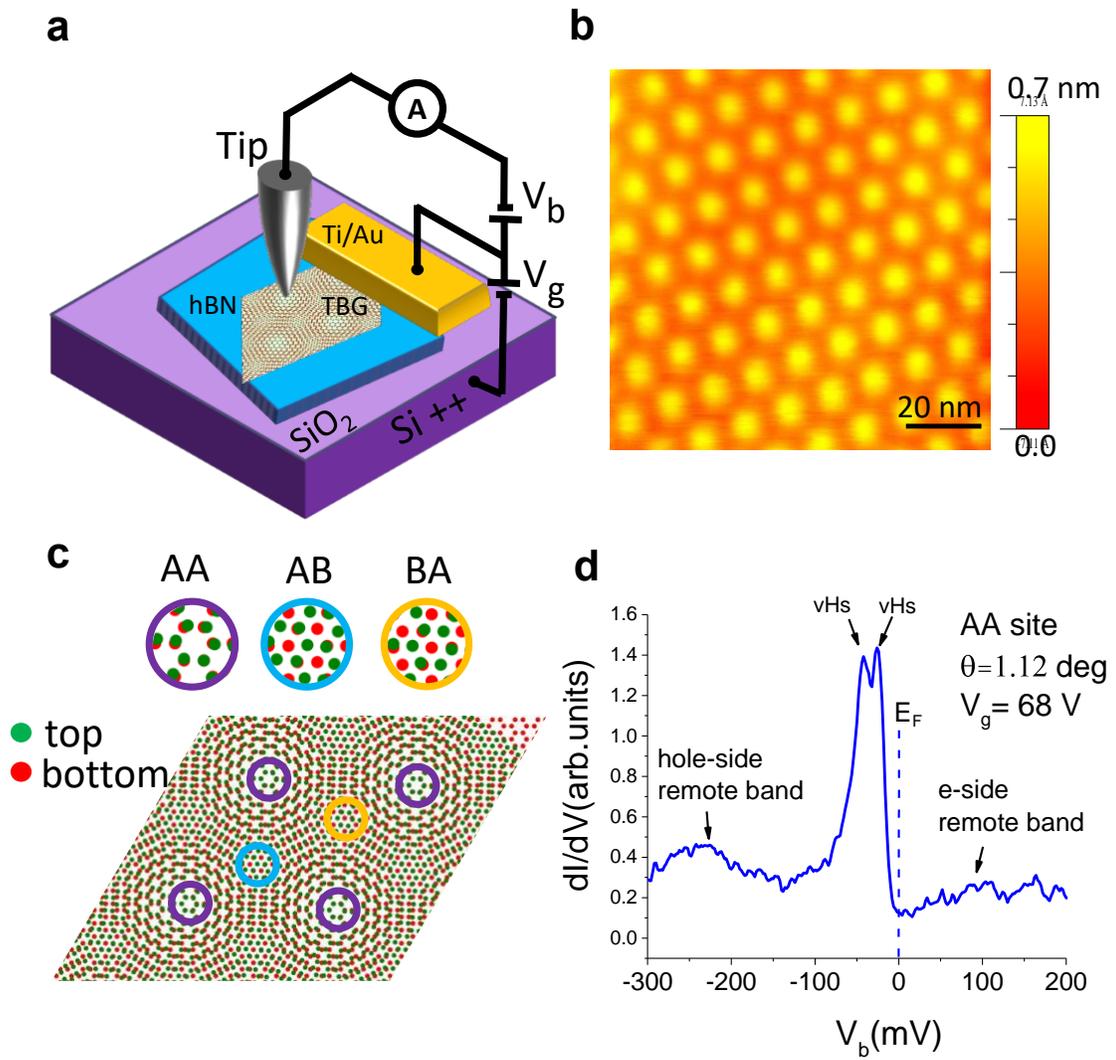

**a**

Tip

A

V_b

V_g

Ti/Au

hBN    TBG

SiO_2

Si ++

**b**

0.7 nm

20 nm

0.0

**c**

AA    AB    BA

● top
● bottom

**d**

AA site
θ=1.12 deg
V_g= 68 V

vHs  vHs

E_F

hole-side
remote band

e-side
remote band

$dI/dV$(arb.units)

$V_b$(mV)

Figure 2

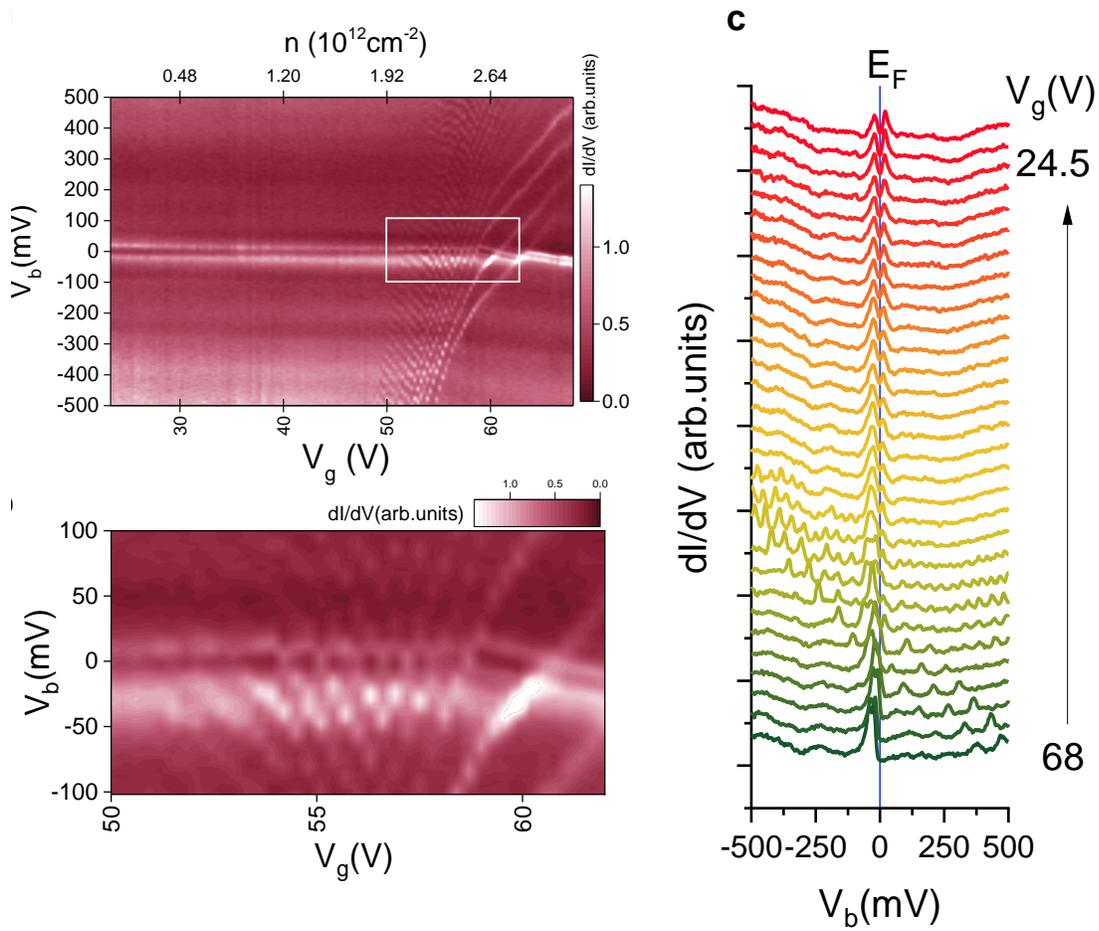

Figure 3

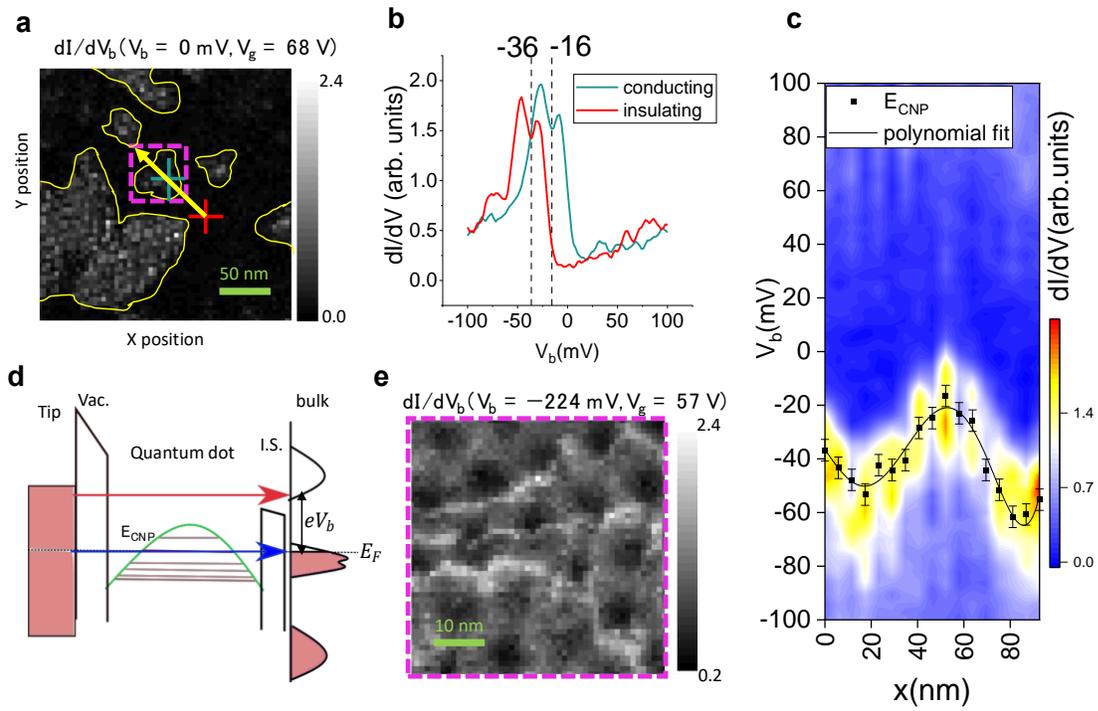

**a** dI/dV$_b$(V$_b$ = 0 mV, V$_g$ = 68 V)

Y position

X position

50 nm

**b**

-36  -16

dI/dV (arb. units)

conducting
insulating

V$_b$(mV)

**c**

E$_{CNP}$
polynomial fit

V$_b$(mV)

dI/dV(arb.units)

x(nm)

**d**

Tip   Vac.   Quantum dot   bulk

I.S.

E$_{CNP}$   eV$_b$   E$_F$

**e** dI/dV$_b$(V$_b$ = −224 mV, V$_g$ = 57 V)

10 nm

**Figure 4**

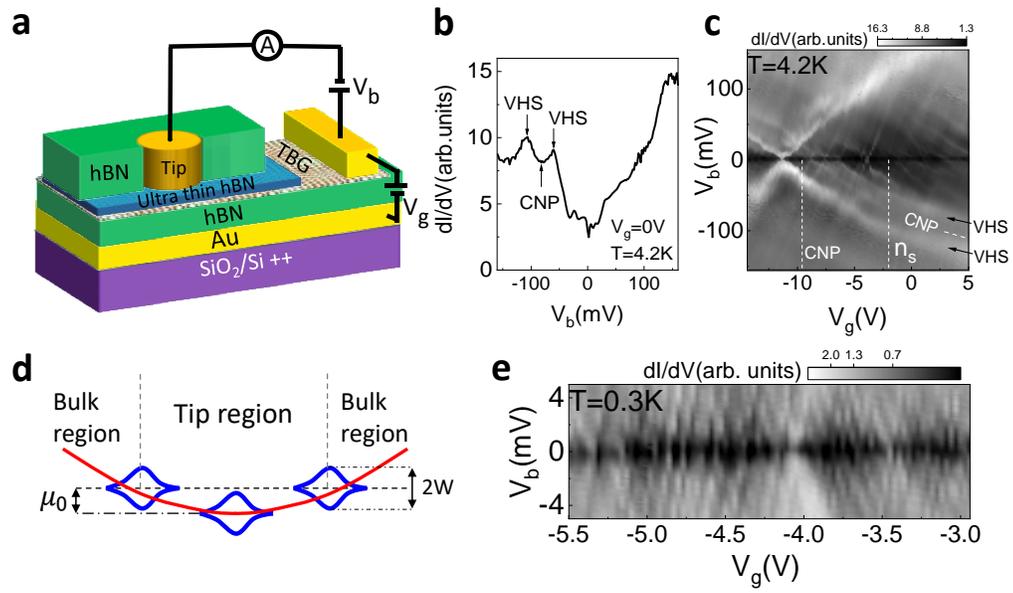

**Figure captions**

**Fig. 1: STM/ STS on Magic-angle twisted bilayer Graphene.**

(a) Schematic of the STM device and measurement setup. The TBG sample was made using the 'twist and stack' method. The Tungsten tip is kept grounded while a bias voltage $V_b$ is applied to the sample. A backgate voltage $V_g$ is applied between the p-doped Silicon backgate and the sample. All measurements were taken at 4.9 K. (b) STM topography measured at a bias of -300 mV and tunneling current of -30 pA. The bright yellow spots which form a triangular lattice are the AA stacking regions. The surrounding darker orange regions have approximately AB and BA stacking. The moiré wavelength was measured from STM topography to be 12.6 nm which corresponds to a twist angle of 1.12°. (c) An illustration of the various stacking arrangements. The green and red dots represent the carbon atoms in the top and bottom Graphene layer respectively. (d) shows a typical $dI/dV_b$ spectrum measured at the center of an AA region at $V_g$=68V when the flat bands are completely full. The van Hove singularities (VHS) and the remote bands on either side of the flat band are labelled. Gap-like dips in the $dI/dV_b$ signal separate the flat bands from the remote bands. The Fermi level $E_F$ of sample is shown with a dashed line at $V_b = 0\ mV$.

**Fig. 2: $dI/dV_b(V_b,\ V_g)$ maps at an AA region.**

(a) Shows the $dI/dV_b(V_b,\ V_g)$ maps at an AA region. At high gate voltages, the flat bands appear as two bright lines below the $E_F$ at $V_b = 0\ mV$, indicating that they are fully filled. As the gate voltage is reduced the $E_F$ intersects the flat bands and they begin to empty. Since the density of states in the flat bands is very high the bands, seen as two horizontal bright lines, look pinned to $E_F$ as they slowly empty. Note that the lowest applied gate voltage of 23.3 V is not enough to fully empty the bands. In addition to these expected spectroscopic features, a series of oblique lines can be seen in a narrow gate range of 50 V-68 V. When these lines intersect the $E_F$, a series of faint coulomb diamonds are seen. These additional spectroscopic features indicate the presence of a quantum dot in the system.

(b) is a zoomed-in view of the region marked by the white rectangle in panel (a). (c) Some selected $dI/dV_b$ spectra from the gate dependence map which show the confinement peaks in the spectra only when the Fermi level is near the band edge. The evolution of the individual confinement peaks with gate voltage appears as the oblique lines in (a) can also be seen.

**Fig. 3: Evidence of local doping variation in the sample.**

(a) shows a $dI/dV_b(x, y, V_b = 0\ mV, V_g = 68\ V)$ spatial map of a $246 \times 246\ nm^2$ region (scale bar 50 nm). Several bright regions, indicating a high density of states, which are surrounded by dark regions, indicating a low density of states are visible. The thin yellow lines are a guide for the eyes. (b) shows individual $dI/dV_b$ spectra measured at the positions marked by the red and green crosses in (a). A lateral shift of the two spectra with respect to the $E_F$ indicates a difference in the local chemical potential. The approximate location of the charge neutrality point taken as the dip between the two VHS is also labelled. (c) is a line-cut across the conducting island in (a) in a direction indicated by the yellow arrow. The variation of the local charge neutrality point energy ($E_{CNP}$) as a function of position can be seen which indicates local doping variation. The black line is a 5th order polynomial fit. Error bars represent uncertainty in the determination of CNP due to finite energy resolution. The bright conducting regions surrounded by the darker insulating regions act like quantum dots. (d) A Tunneling diagram of the system. There are two tunneling barriers: the large vacuum barrier between the tip and the dot and the shorter barrier between the dot and the bulk of the sample. The quantum dot is defined by a potential well which has the shape indicated by the local charge neutrality point. Confined states within this potential well are indicated by horizontal lines. There are low energy electron tunneling events from the tip into the confined states of the quantum dot and then into the bulk of the sample (blue arrow). There also exist direct tunneling from the tip to the higher energy bands of the bulk of the sample (red arrow). (e) shows a high resolution $dI/dV_b(x, y, V_b = -224\ mV, V_g = 57\ V)$ spatial map of the conducting island indicated by the magenta square in (a). The three bright wave-fronts are three single electron charging states confined in the bright island. The shape of the bright wave fronts approximately matches the shape of the bright island.

**Fig. 4: Carrier confinement in a TBG planar Tunneling device. a** Schematic of the planar tunneling device. An ultra-thin hBN acts as a tunneling barrier between the metallic tip and the TBG. **b** Tunneling spectrum measured at 0 V backgate voltage showing the flat bands. The sample is highly n-doped because of the work function difference between the tunneling electrode and the TBG. **c** gate dependence map of the STS. Flat bands can be seen as two bright parallel lines at negative bias. The other features are a result of confinement. **d** Schematic of the position dependence of the charge neutrality point. $\mu_0$ is the shift in the charge neutrality point under the tip compared to the bulk region. **e** High resolution $dI/dV_b$ gate dependence maps measured at 0.3 K showing coulomb diamonds, a result of electron-confinement.

# Supplementary Information for 'Flat band carrier confinement in magic-angle twisted bilayer graphene'

Nikhil Tilak[1], Xinyuan Lai[1], Shuang Wu[1], Zhenyuan Zhang[1],
Mingyu Xu[2,3], Raquel de Almeida Ribeiro[2,3], Paul C Canfield[2,3]
and Eva Y. Andrei[1*]

1 *Department of Physics and Astronomy, Rutgers, The State University of New Jersey, 136*

*Frelinghuysen Rd, Piscataway, NJ 08854*

2 *Ames Laboratory, U.S. Department of Energy, Ames, Iowa 50011, USA*

3 *Department of Physics and Astronomy, Iowa State University, Ames, Iowa 50011, USA.*

Corresponding author email: eandrei@physics.rutgers.edu

# Supplementary Note 1: Device micrographs

(a)

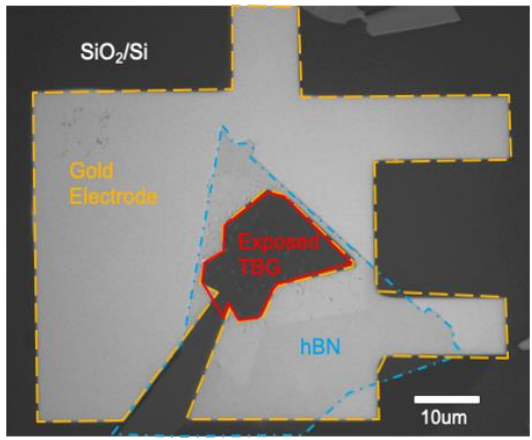

(b)

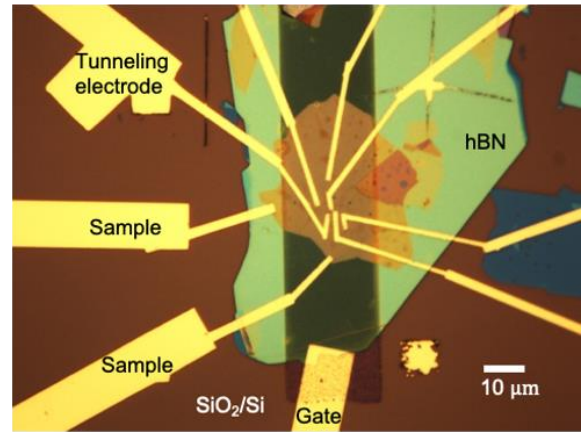

Supplementary Figure 1 (a) Optical micrograph (grayscale) of the STM device. The Gold electrode (yellow dashed lines), TBG region (red solid line) and hBN (cyan dashed line) are marked. (b) Optical micrograph of the planar tunneling device.

# Supplementary Note 2: Waterfall plots of dI/dV gate dependence

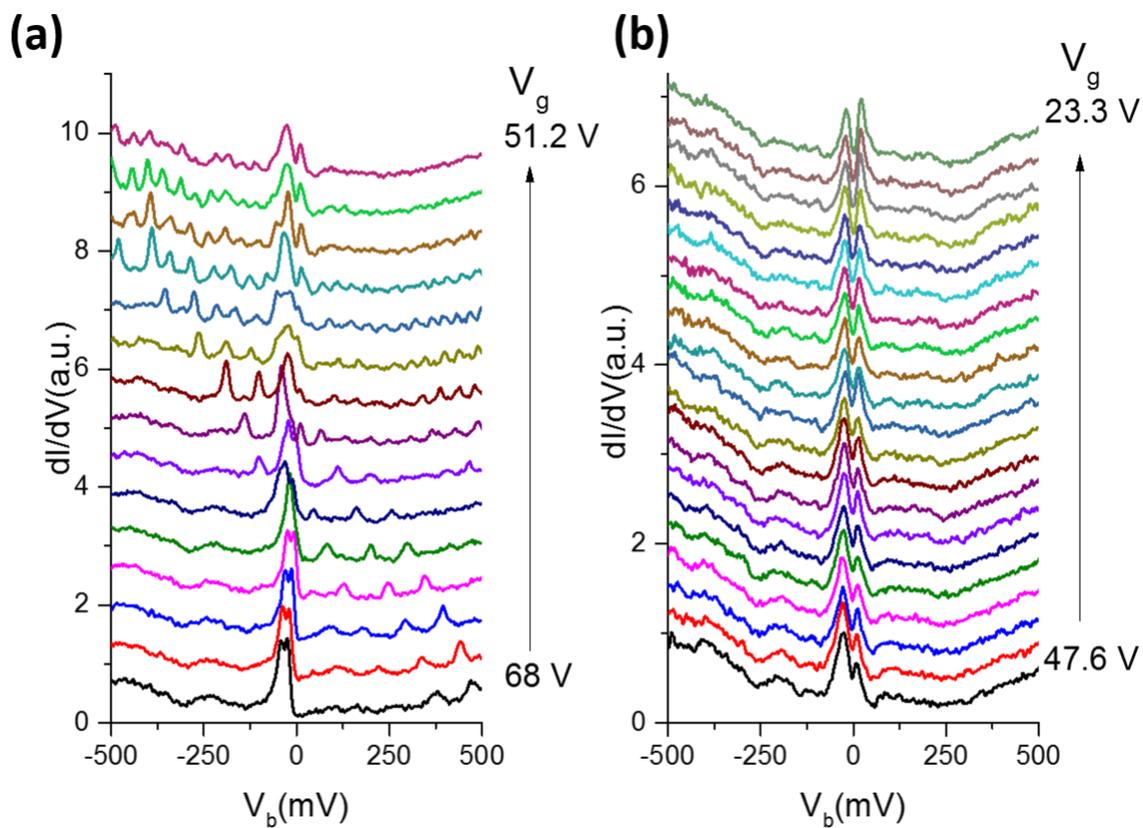

Supplementary Figure 2 dI/dV evolution with gate voltage from 68 V to 51.2 V (a) and from 47.6 V to 23.3 V (b) respectively

(same data as main Fig 2). Quantum dot peaks are no longer visible in the bias range when the gate voltage is below about 50 V.

# Supplementary Note 3: Coulomb diamond analysis

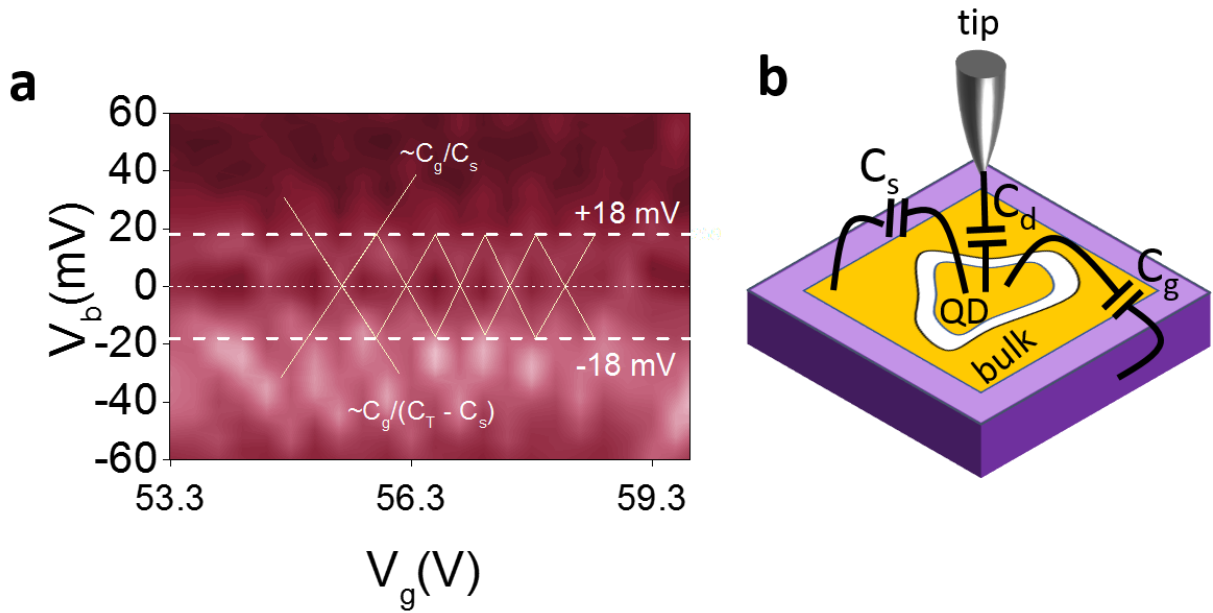

Supplementary Figure 3 (a) A magnified view of the $dI/dV_b(V_b, V_g)$ map from Main Fig. 2a which shows the Coulomb diamonds near the low bias range. The yellow lines are guides for the eyes. The slopes of the coulomb diamond have also been labelled. $C_g$, $C_s$ and $C_d$ are the capacitances between gate-dot, bulk-dot and tip-dot respectively as illustrated in panel (b). The total capacitance $C_T = C_g + C_s + C_d$.

Each quantum dot is capacitively coupled to the backgate, the tip and the nearby conducting regions of the sample. Tunneling from the tip into the dot and then from the dot to the nearby conducting regions of the sample requires a precise alignment of energy levels of the dot with those of the tip and the sample (Main Fig. 3d). This alignment depends on all three capacitances, the sample bias and the gate voltage as discussed below.

First let's consider the case where the bias voltage is smaller than the potential barrier surrounding the quantum dot. In this case the only way in which electrons can tunnel from the tip into the bulk of the sample is by first tunneling into the quantum dot. As the backgate voltage is increased (decreased), the energy states in the quantum dot are lowered (raised). When an energy level of the quantum dot aligns with the sample Fermi level, electrons can tunnel into the quantum dot from the tip and out of the quantum dot into the bulk of the sample. This is observed as a sharp increase in the tunneling current. Because of Coulomb repulsion, the next electron needs an additional energy ($E_{add}$) to tunnel into the dot. The additional energy is, in general, the sum of the charging energy ($E_C$) and the orbital spacing ($\Delta$). The orbitals have some degeneracy. The energy $\Delta$ is only needed once all the states in a particular orbital have been filled and the electron is added to the next orbital. The charging energy depends on the total capacitance of the quantum dot ($C_T = C_g + C_s + C_d$) via $E_C = e^2/C_T$. The orbital spacing $\Delta$ depends on the size of the dot and the details of the confining potential. The next higher energy state becomes accessible when the gate voltage is further increased by a value $\Delta V_g = e/C_g$.

At larger sample bias ($V_b$), the tip can also gate the confined states in the dot. A negative (positive) sample-bias is equivalent to a positive (negative) tip-bias and has the same effect as a positive (negative) backgate voltage. The same quantum dot state can be aligned with the tip/sample Fermi level for a continuous set of linear combinations of sample bias and backgate voltage. This is the origin of the series of oblique lines seen in the gate dependence maps (Main Fig. 2). The slope of these lines contains information about the capacitances between the tip-dot ($C_d$) and the dot-sample ($C_s$). To further study the role played by the tip-dot capacitance we recorded the tunneling set-point dependence of the spectra (see Supplementary Figure 7).

By analyzing the Coulomb diamonds as outlined in Supplementary Figure 3a, we measured the charging energy of our quantum dot, $E_C = (18 \pm 1.4)$ meV, which gives $C_T = (8.8 \pm 0.7)$ aF. The capacitance between the dot and the back gate in our sample is $C_g = 2.4 \pm 0.6$ aF. Modelling the quantum dot as a circular metallic disc, we estimate the size of the quantum dot as $61 \pm 4.8$ nm. This matches reasonably well with the size (~50-60 nm) of the conducting region where the spectrum in Fig. 2a was collected. It is worth noting that the apparent size of the quantum dot as well as the height of the tunneling barrier depend on the applied gate voltage. Approximating the quantum dot as a flat disc with diameter 60 nm and using a carrier effective mass of $m^* = 0.05 m_e$ from Bernal Bilayer Graphene[1], we found that $\Delta \approx 2$ meV which is negligible compared to the measured charging energy of 18 meV. $\Delta$ can therefore safely be neglected.

By analyzing the slopes of the charging lines we calculated the ratio of the three capacitances as $C_g : C_s : C_d :: 1 : 37 : 36$. This implies that although $V_b$ is typically small compared to $V_g$, the tip is very efficient at gating because of its proximity to the dot. This explains the sign of the slope of the oblique charging lines in the gate dependence map (Main Fig. 2a).

# Supplementary Note 4: Additional gate dependence maps on AA and AB sites

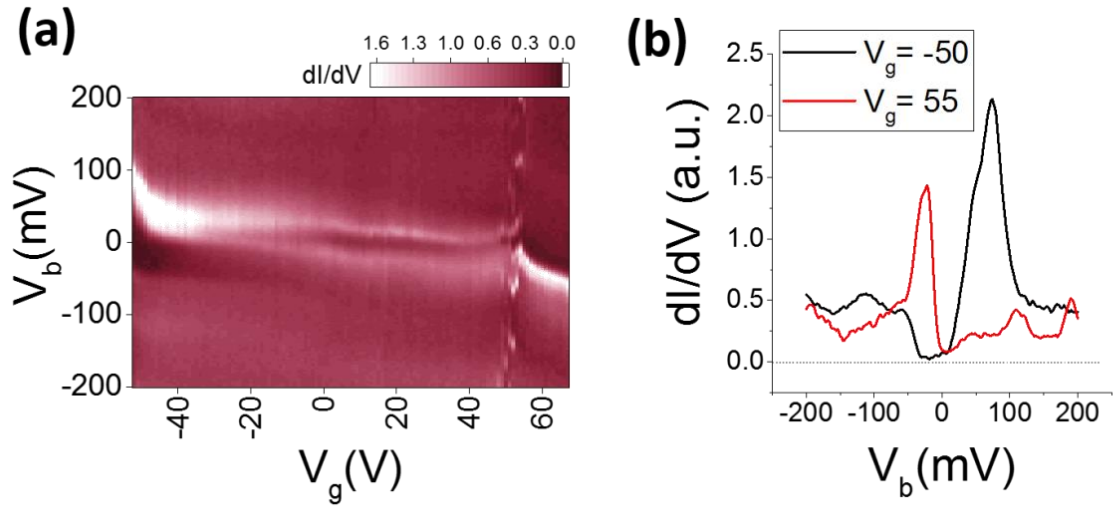

Supplementary Figure 4 (a) A wide gate-range $dI/dV_b(V_b, V_g)$ map at an AA site located in the same region as Main Fig. 3a. It takes about 80 V of gate voltage to go from fully filled bans to fully empty bands based on the capacitance between the backgate and sample. (b) Individual $dI/dV_b$ spectra when the bands are fully filled ($V_g = 55\,V$) and fully empty ($V_g = -50\,V$). The superlattice gaps can be seen near the fermi level in both curves but the gap on the other side of the flat bands is not clearly visible because of a large background signal at higher biases.

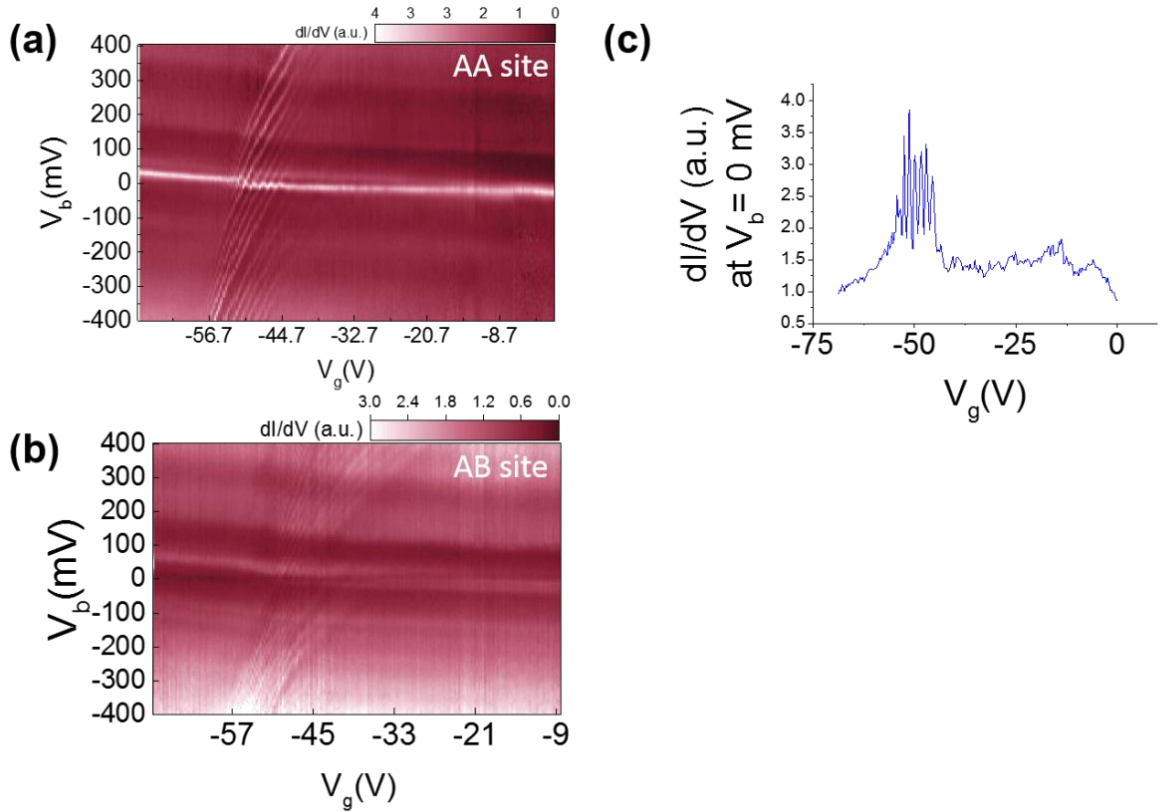

Supplementary Figure 5 Charging peaks seen on the bottom edge of the hole side flat band in a different region in the sample measured at an AA site (a) and at an AB/BA site (b). The zero bias conductance measured from the data in (a) is shown in (c). The single electron charging events at zero bias appear as sharp peaks.

Although the results in the main figures demonstrate confinement near the electron-side band-edge, there is nothing special about the electron-side band. Confinement can also exist in the hole side band edge. We measured confinement states near the hole side band edge at a different position in the sample. results are shown in Supplementary Figure 5b.

# Supplementary Note 5: High resolution dI/dV spatial mapping

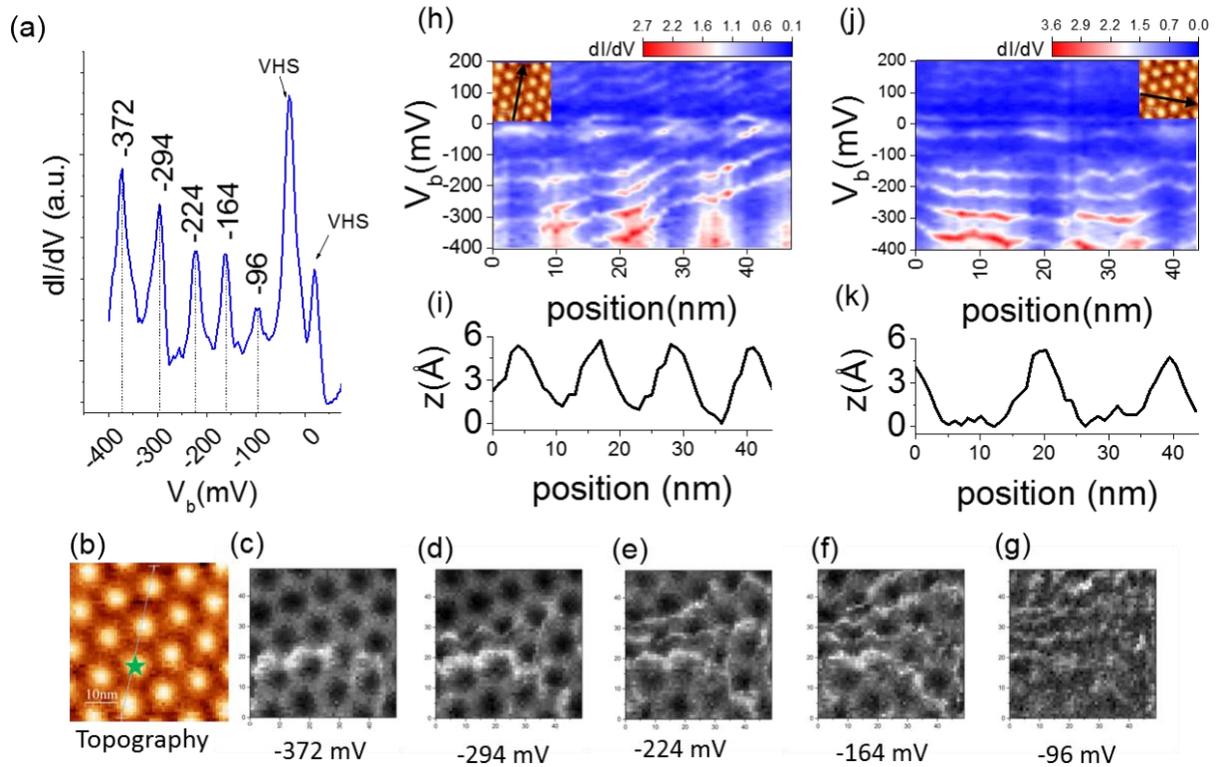

Supplementary Figure 6 dI/dV spatial maps at $V_g$ = 57 V. (a) shows an STS spectrum measured at position marked by the green star in topography (b). Charging peaks corresponding to the Quantum dot states are seen at -372 mV, -294 mV, -224 mV, -164 mV and -96 mV respectively. The dI/dV maps corresponding to these energies are shown in figures (c)-(g). Bright wavefronts corresponding to each QD state are seen in these maps. Figure (h) maps the position dependence of dI/dV at various points along the direction labelled by the arrow in the inset. Figure (i) shows the corresponding height variation along the same direction. The Quantum dot peaks in (h) show a general trend of shifting higher in energy (less negative) as the tip moves outwards towards the edge of the Quantum dot. This is an indication that the tip-dot capacitance reduced as the tip moves towards the edge of the Quantum dot. This reduces the tip-gating efficiency and thus the same QD state appears at higher bias voltages near the edge of the dot. The change in energy briefly slows down when the tip is scanning an AA site as marked by the dotted lines. This can be attributed again to the decrease in the gating efficiency of the tip as the tip-dot distance increases in the z direction while scanning the AA site. Figures (j) and (k) are similar to (h) and (i) except the direction is as indicated by the inset which is perpendicular to the direction for (h) and (i). Along this direction the Quantum dot peaks appear to have nearly constant energy. This is to be expected since this direction is along the QD wavefronts as seen in the dI/dV maps.

To further study the Quantum dot at a smaller scale we performed dI/dV mapping in the region marked by the magenta square in Main Fig. 3a at a gate voltage of 57 V. Supplementary Figure 6a shows the dI/dV spectrum measured at the position labelled by the green star in Supplementary Figure 6b. In addition to the flat bands on either side of 0 mV bias, several peaks corresponding to the charging of QD states are visible in the spectrum. In Supplementary Figure 6(c-g) the dI/dV maps at these energies are shown. In the maps the charging states are seen as bright wavefronts. At higher energies all the wavefronts with lower energies are also visible. As the energy is increased from negative values towards 0, the wavefronts appear to move outward toward the edge of the QD. This can be explained by considering the decrease in the tip-dot capacitance as the tip moves from the center of the dot towards the edge. As the tip-dot capacitance decreases, the tip gating efficiency goes down. This means that we need a larger (less negative) bias in order to reach the same QD energy level.

This effect can be seen more directly by mapping the position dependence of the QD energy levels along a direction which points towards the edge of the QD (Supplementary Figure 6h) and a direction orthogonal to it (Supplementary Figure 6j). In Supplementary Figure 6h, the QD states can be seen moving up in energy as the tip moves outward towards the edge of the dot. Interestingly the increase in the QD peak energies seems to slow down when the tip is close to the AA regions. This again can be explained by considering the reduction in the tip-gating efficiency as the tip-sample height is greater by about 0.5 nm (Supplementary Figure 6i), when the tip is near the top of the AA sites owing to the larger LDOS there.

In contrast, in Supplementary Figure 6j, the energy of the QD states doesn't seem to change much. This is because the tip-dot capacitance doesn't change appreciably when the tip moves along this direction. Notice that this direction is roughly tangential to the wavefronts.

# Supplementary Note 6: dI/dV set-point dependence

**(a)**

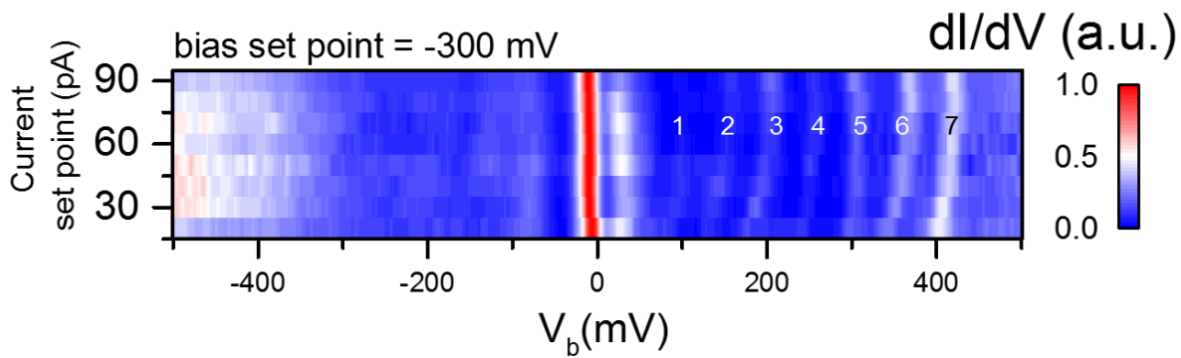

**(b)**

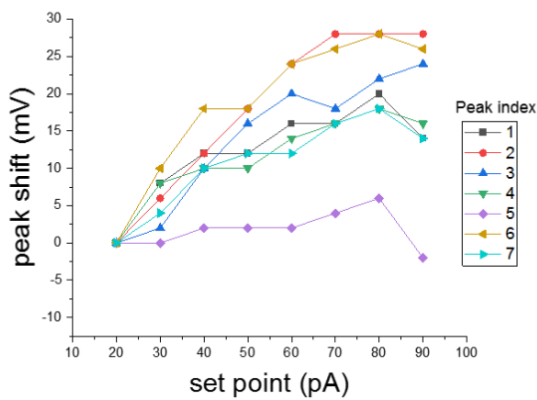

**(c)**

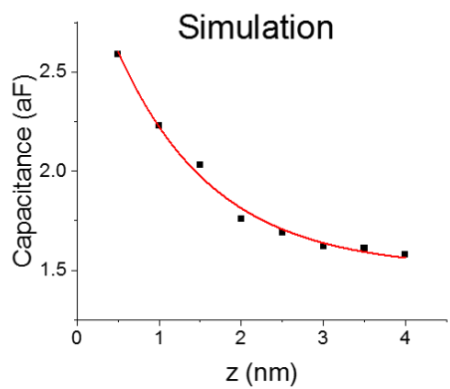

Supplementary Figure 7 Set point dependence of the confinement states. (a) is a set point dependence dI/dV map. current set points were varied between 20 pA and 90 pA while maintaining an initial bias set-point of -300 mV. The bright lines at positive biases represent confined states labelled by numbers 1 -7. As the set point in increased the energies of the states systematically increase. (b) plots the shift in the energy of each of the 7 states as the set point is increased from 20 to 90 pA. The states shift upwards in energy by about 20 meV. Interestingly state 5 doesn't change significantly in energy. The capacitance between the tip and the sample should increase as the tip is brought closer to the sample. We simulated our tip as a sphere and the sample as a circular disc. The calculated Capacitance between the tip and the sample is plotted in (c) as a function of tip-sample distance. The capacitance indeed increases non linearly as the tip-sample distance is reduced.

Another way of changing the tip-dot capacitance is by changing the initial set point at which the dI/dV spectrum is measured. To this end, we positioned the tip on top of an AA site and measured the dI/dV spectra at initial current set points of 20 pA to 90 pA in increments of 10 pA keeping the initial bias set point constant at -300 mV. In Supplementary Figure 5a we plot the spectra as a map. The bright lines, most clearly seen at positive bias, are QD states. They have been labelled 1-7. Observe that as the set point is increased, the states move to higher energies. This increase in energy is plotted in Supplementary Fig. 5b. Almost all the states increase in energy by about 20 meV. Interestingly state 5 doesn't seem to move much. We have been unable to explain this result.

As a sanity check we simulated the system as a spherical tip with radius 10 nm and the Quantum dot as a circular disc of radius 100 nm and calculated the capacitance between them for different tip-sample distances. The results are shown in Supplementary Figure 5c. Each data point represents a simulation performed with a different tip-sample distance. The capacitance increases non-linearly as the tip-sample distance is reduced.

# Supplementary Note 7: Carrier density variation estimation in the STM device

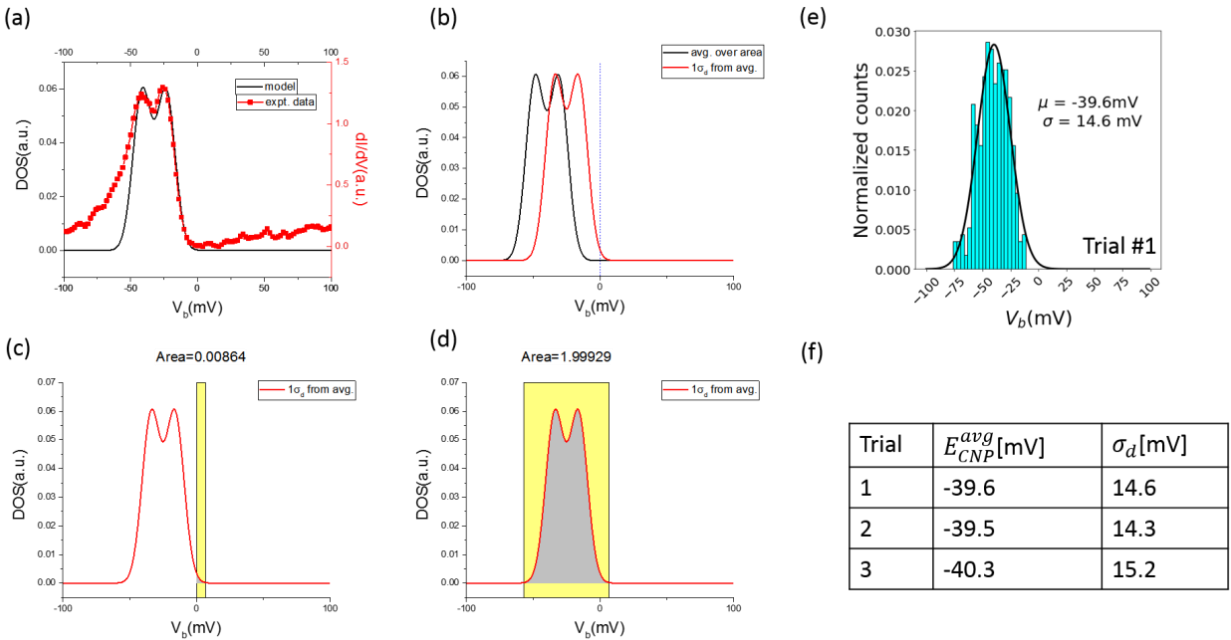

Supplementary Figure 8: Carrier disorder estimation (a) The flat bands are modelled as two gaussians with centers separated by 18 mV and FWHM of 16 mV each. The model (black curve) is superimposed on an experimental dI/dV curve (red) to show the goodness of fit. The model underestimates the density of states at the lower band edge, most likely because of a non-linear background signal. (b) The model is plotted according to the average CNP position in the map (black) and 1 std. dev. shifted to the right (red). (c) the gray shaded area under the DOS curve integrated from the fermi level (0 mV) to the upper band edge (6.9 mV) represents empty states. (d) the gray shaded area under the DOS curve integrated from the lower band edge (-57.1 mV) to the upper band edge (6.9 mV) represents the total states within the flat band. (e) An example of the charge neutrality point variation in the dI/dV spatial map (main Fig 3) displayed as a histogram. 360 points were randomly sampled from the dI/dV map and the CNP energy at those points was noted. The black line is a gaussian $(\mu, \sigma)$ curve fitted to the histogram. (f) shows the tabulated results for 3 such trials.

In previous STM experiments on Monolayer Graphene (MLG) supported by a SiO$_2$, hBN or MoS$_2$ substrate, the carrier density variation was estimated by measuring the average variation of the Dirac point energy ($E_d$) and translating that to a carrier density variation using the well-known formula for the $E_d(n)$ for MLG: $E_d^2 = \hbar^2 v_F^2 \pi \, n$ .

In TBG, the exact analytical expression for the dispersion is not known. Therefore, the simple approach used for MLG cannot be directly used in the case of TBG. In what follows, we attempt to get an estimate for the carrier density disorder in our sample.

First, to quantify the variation of the charge neutrality point ($E_{CNP}$) in the sample we used the following procedure-

1. A large dI/dV spatial mapping was acquired at a gate voltage $V_{g0} = 68\,V$ (Main Fig 3a) and the energy of the charge neutrality point ($E_{CNP}$) in the dI/dV spectra taken at various positions in the mapping was recorded. This was done with the help of a Python script which randomly samples 360 points in the 3600 point dI/dV map. Then the STS at each of these points is extracted from the data. The dip in the dI/dV spectrum between the two VHS is manually assigned as the CNP. The actual CNP position may not be exactly at the mid-point between the VHSs but it is a reasonable approximation for the purposes of this calculation.

2. The charge neutrality point energy distribution is plotted as a histogram and a gaussian was fitted to it. The standard deviation ($\sigma_d$) of the fitted gaussian is taken as a measure of the chemical potential variation in the sample.

3. This procedure of randomly sampling 10% of the points in the grid was repeated 3 times in order to get a better sense of the average variation.

4. The average $\sigma_d$ using this procedure is $\sigma_d = (14.7 \pm 0.5)$ mV and the average position of the CNP was $E_{CNP}^{avg} = -39.8 \pm 0.4$ mV. The error bars represent one standard deviation calculated from the 3 trials.

After this procedure we estimated the carrier density disorder ($n_d$) as follows-

1. The separation of the VHS and the FWHM were recorded for many points from the STS map. This yielded an average VHS separation of 18 mV and the FWHM of each band as $w = 16\ mV$.
2. The DOS of the flat bands was modelled as the sum of two gaussians with the centers separated by 18 mV and with FWHM $w = 16\ mV$. The height of the gaussians was determined by the area under the curve which was arbitrarily set to 1. The exact value of the area does not affect the following calculations.
3. The band edges are taken as $E_{CNP}^{avg} \pm 2w$. For the measured average value $E_{CNP}^{avg} = -39.8$ mV average position the band edges are located at (-71.8 mV, -7.8 mV).
4. This implies that the $E_F = 0\ mV$ is located inside the gap between the upper flat band and the upper remote band i.e. the flat bands are fully filled. Now consider a spatial position in the sample where the CNP position is shifted with respect to the average value by 1 $\sigma_d$ due to doping disorder. The upper band edge will then be shifted to $\delta = E_{CNP}^{avg} + 2w + \sigma_d$ mV. The measured value of $\sigma_d = 14.7\ mV$ places the band edge at $6.9 \pm 0.9$ mV where the error bar represents the uncertainty in the values of $\sigma_d$ and $E_{CNP}^{avg}$.
5. The states from the $E_F$ to the band edge, i.e., 0 mV to 6.9±0.9 mV are empty. Integrating the simulated DOS in this range gives an area of $\int_0^{6.9 \pm 0.9} DOS(E)dE = 0.009 \pm 0.003$ units. Comparing this to the integration of the entire bands $\int_{-57.1}^{6.9} DOS(E)dE = 1.999$ gives an empty band fraction of $\frac{0.009 \pm 0.003}{1.999} = 0.0045 \pm 0.0015$. Since the flat bands at the measured twist angle of 1.12 deg can accommodate $2n_s = 5.82 \times 10^{12}/cm^2$ carriers, a 0.0045±0.0015 fraction of that represents $(2.62 \pm 0.87) \times 10^{10}/cm^2$ carriers. This gives an estimate of the carrier density variation in the sample.

# Supplementary Note 8: Quantum dots in STS experiments from trivial origins

There are various scenarios where resonant states can appear in STS. One trivial possibility is that a weakly conducting impurity can be attached to the STM tip near the tunneling junction. This impurity can act like a Quantum dot and produce sharp peaks in STS whenever the sample or tip Fermi level aligns with its confinement states. Since such a Quantum dot is always surrounded by a tunneling barrier which is independent of the gate voltage, these peaks show up at all gate voltages. Main Figure 2c shows dI/dV spectra gathered at different gate voltages. It can be seen that the resonant states exist only for a small range of gate voltage. Furthermore, we reformed the tip apex by applying bias pulses (3-6 VDC for 200 ms) while scanning the gold electrode. This procedure re-sculpts the very apex of the tip near the tunneling junction and is almost equivalent to changing tips. This procedure was repeated until sharp topographic scans were obtained. Then dI/dV spectra were measured on the Gold electrode to ensure that the spectrum was flat in the energy range of interest ($\pm$ 500 mV). The resonant peaks in the TBG region were found to be reproducible before and after such tip cleaning. Therefore, we can safely rule out the possibility of a nanoparticle loosely stuck to the tip as a source of these peaks.

Another possibility is that the tip is invasive and creates local p-n junctions[2] or causes band bending locally[3,4] in order to produce confinement. We can safely eliminate this possibility because the Quantum dot states occur in specific locations in our sample unlike tip-induced states which should not be position dependent.

# Supplementary Note 9: Additional Planar Tunneling Spectra at 77K

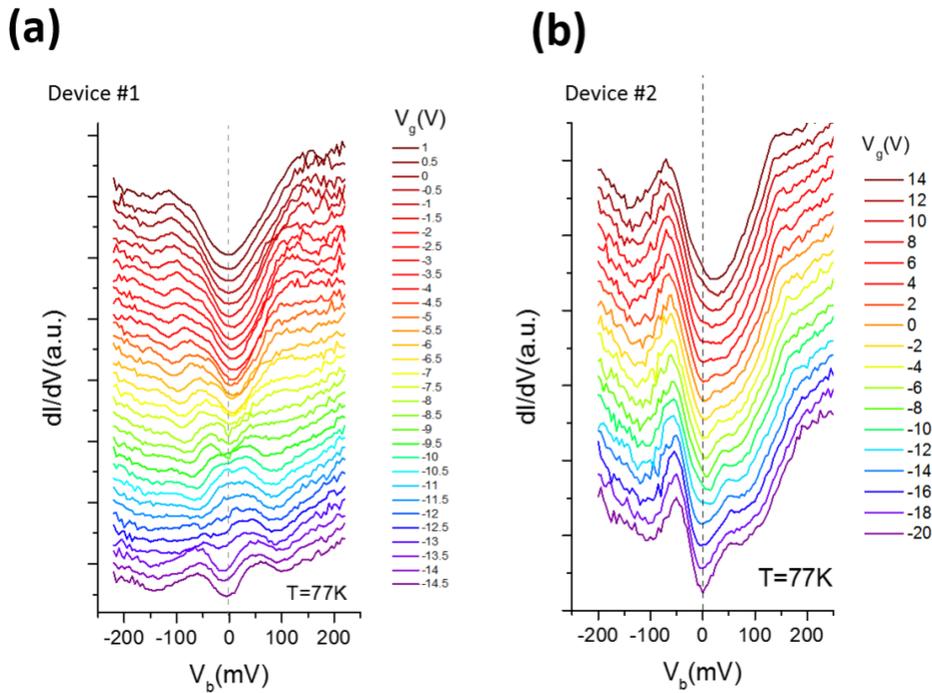

Supplementary Figure 9 dI/dV ($V_b$, $V_g$) of device #1, #2 at 77K. Device #2 shows a similar doping effect as device #1 which was presented in the main paper, confirming the reproducibility of our results.

# Supplementary Note 10: Estimation of the twist angle of PTJ device from gate dependence

As a check we also estimate the twist angle from the gate voltage needed to completely fill the flat bands using equations (2) and (3) as described in the main paper. At $V_g = 0$, the average of the energies of the upper and lower

VHS approximately corresponds to the energy of the CNP of the sample (~-80 mV). This indicates that the sample is electron-doped, presumably due to the work function difference between the tunneling electrode and the TBG. Extrapolating the CNP position to the Fermi level ($V_b = 0$) indicates that the sample is charge neutral at a $V_g = -9.7$ V labeled by the dashed white line in Fig 4c. Since our gating ability is limited to $\pm 14.5\ V$ by the breakdown voltage of the bottom hBN dielectric layer we were unable to completely empty the flat bands. Nonetheless, since the flat bands start emptying at $V_g \approx -(2 \pm 1)$ V and are filled to the CNP at $V_g \approx -9.7$ V we can estimate that it takes about $15.4 \pm 2$ V of gate voltage to completely empty the bands. Inserting $(V_g - V_{g0}) = 15.4$ V, $d_1 = 70$ nm (thickness of bottom hBN) and $d_2 = 0$ in equation (3) yields a charge density of $(4.9 \pm 0.6) \times 10^{12} \mathrm{cm}^{-2}$. Since this charge density corresponds to $2n_S$, equation (2) gives a twist angle of $\sim (1.0 \pm 0.1)^\circ$.